\documentclass[twocolumn,superscriptaddress,aps,longbibliography,showpacs,preprintnumbers,amsmath,amssymb,floatfix]{revtex4-1}
\usepackage{titlesec}
\usepackage{amsmath}
\usepackage{graphicx}
\usepackage[ansinew]{inputenc}
\usepackage{array}
\usepackage{color}
\usepackage{amsxtra}
\usepackage{amstext}
\usepackage{amssymb}
\usepackage{latexsym}
\usepackage{gensymb}
\usepackage{dsfont}
\usepackage{braket}
\usepackage[caption=false]{subfig}
\usepackage[makeroom]{cancel}

\usepackage{dcolumn}
\usepackage{bm}
\usepackage{bbm}
\usepackage{braket}
\usepackage{mathrsfs}
\usepackage{amstext}
\usepackage{euscript}
\usepackage{txfonts}

\usepackage[unicode=true,
 bookmarks=false,
 breaklinks=false,pdfborder={0 0 1},backref=false,colorlinks=true,citecolor=blue]
 {hyperref}

\makeatletter
\@ifundefined{textcolor}{}
{%
 \definecolor{BLACK}{gray}{0}
 \definecolor{WHITE}{gray}{1}
 \definecolor{RED}{rgb}{1,0,0}
 \definecolor{GREEN}{rgb}{0,1,0}
 \definecolor{BLUE}{rgb}{0,0,1}
 \definecolor{CYAN}{cmyk}{1,0,0,0}
 \definecolor{MAGENTA}{cmyk}{0,1,0,0}
 \definecolor{YELLOW}{cmyk}{0,0,1,0}
}

\makeatletter
\renewcommand*\env@matrix[1][*\c@MaxMatrixCols c]{%
  \hskip -\arraycolsep
  \let\@ifnextchar\new@ifnextchar
  \array{#1}}
\makeatother

\newcommand{\cref}[1]{Ref.\,\cite{#1}}

\makeatother

\begin{document}


\title{Dark-State Optical Potential Barriers with Nanoscale Spacing}
\author{Wenchao Ge}
\affiliation{Institute for Quantum Science and Engineering (IQSE) and Department of Physics and Astronomy, Texas A\&M University, College Station, TX 77843-4242, USA}

\author{M. Suhail Zubairy}
\affiliation{Institute for Quantum Science and Engineering (IQSE) and Department of Physics and Astronomy, Texas A\&M University, College Station, TX 77843-4242, USA}

\date{\today}

 \begin{abstract}
Optical potentials have been a versatile tool for the study of atomic motions and many-body interactions in cold atoms. Recently, 
optical subwavelength single barriers were proposed to enhance the atomic interaction energy scale, which is based on non-adiabatic corrections to Born-Oppenheimer potentials. Here we present a study for creating a new landscape of non-adiabatic potentials---multiple barriers with subwavelength spacing at tens of nanometers. To realize these potentials, spatially rapid-varying dark states of atomic $\Lambda$-configurations are formed by controlling the spatial intensities of the driving lasers. As an application, we show that bound states of very long lifetime on the order of seconds can be realized. Imperfections and experimental realizations of the multiple barriers are also discussed.

\end{abstract}
 \maketitle

\section{Introduction}

Optical potentials have been a very useful tool for manipulation of atomic motions, such as simulation of many-body physics \cite{Bloch:2005aa, Bloch:2008aa}. The typical potentials can be generated using a far-detuned light  through optical dipole force due to a spatially varying AC-Stark shift \cite{Grimm:2000aa}. The potential landscape is determined therefore by the spatial variation of the light intensity. Except operating near the surface \cite{Chang:2009aa, Gullans:2012aa,Romero-Isart:2013aa,Thompson:2013aa,Mitsch:2014aa,Gonzalez-Tudela:2015aa} or using special masks \cite{Brezger:1999aa, Huang:2009aa}, the spatial intensity variation is often limited by the wavelength $\lambda$ of the light. To realize subwavelength resolution in the far field, various approaches have been proposed  \cite{Gardner:1993aa, Qi:2012aa, Yavuz:2009aa, Liao:2010aa,Sun:2011aa,Nascimbene:2015aa,Yi:2008aa,Lundblad:2008aa, Shotter:2008aa,Berman:1998aa, Weitz:2004aa, Salger:2007aa, Ge:2013aa,Sahrai:2005aa,Agarwal:2006aa,Yavuz:2007aa, Gorshkov:2008aa,Kiffner:2008aa,Li:2008aa, Mompart:2009aa,Viscor:2012aa, Miles:2013aa}, such as multi-tone dressing \cite{Yi:2008aa,Lundblad:2008aa, Shotter:2008aa}, multi-photon process \cite{Berman:1998aa, Weitz:2004aa, Salger:2007aa, Ge:2013aa}, and atomic dark states in $\Lambda$-configurations \cite{Sahrai:2005aa,Agarwal:2006aa,Yavuz:2007aa,Gorshkov:2008aa, Kiffner:2008aa,Li:2008aa, Mompart:2009aa,Viscor:2012aa, Miles:2013aa}.

Recently, a novel idea of subwavelength optical potentials was proposed by considering the non-adiabatic corrections \cite{Dum:1996aa} to the spatially varying dark states in the $\Lambda$ configurations \cite{lacki:2016aa,Jendrzejewski:2016aa}. A subwavelength potential barrier arises when the kinetic energy of an atom experiences a rapid change of its internal state in a subwavelength region. This new type of optical potentials is drastically different from the optical dipole potentials as the former is purely quantum mechanical since the potential energy is proportional to $\hbar$. The first experiment of these barriers has been demonstrated \cite{Wang:2018aa}. By using standing-waves,  the potentials can form lattices with subwavelength barriers spaced by $\lambda/2$. With time-dependent engineering of the lattices, smaller spacings between narrow barriers are possible \cite{subhankar2019floquet,lacki2019stroboscopic, Tsui}. However, heating can arise and the dark-state lifetime can be limited due to lattice modulation \cite{lacki2019stroboscopic, Tsui}.

\begin{figure}[t]
\leavevmode\includegraphics[width = 1 \columnwidth]{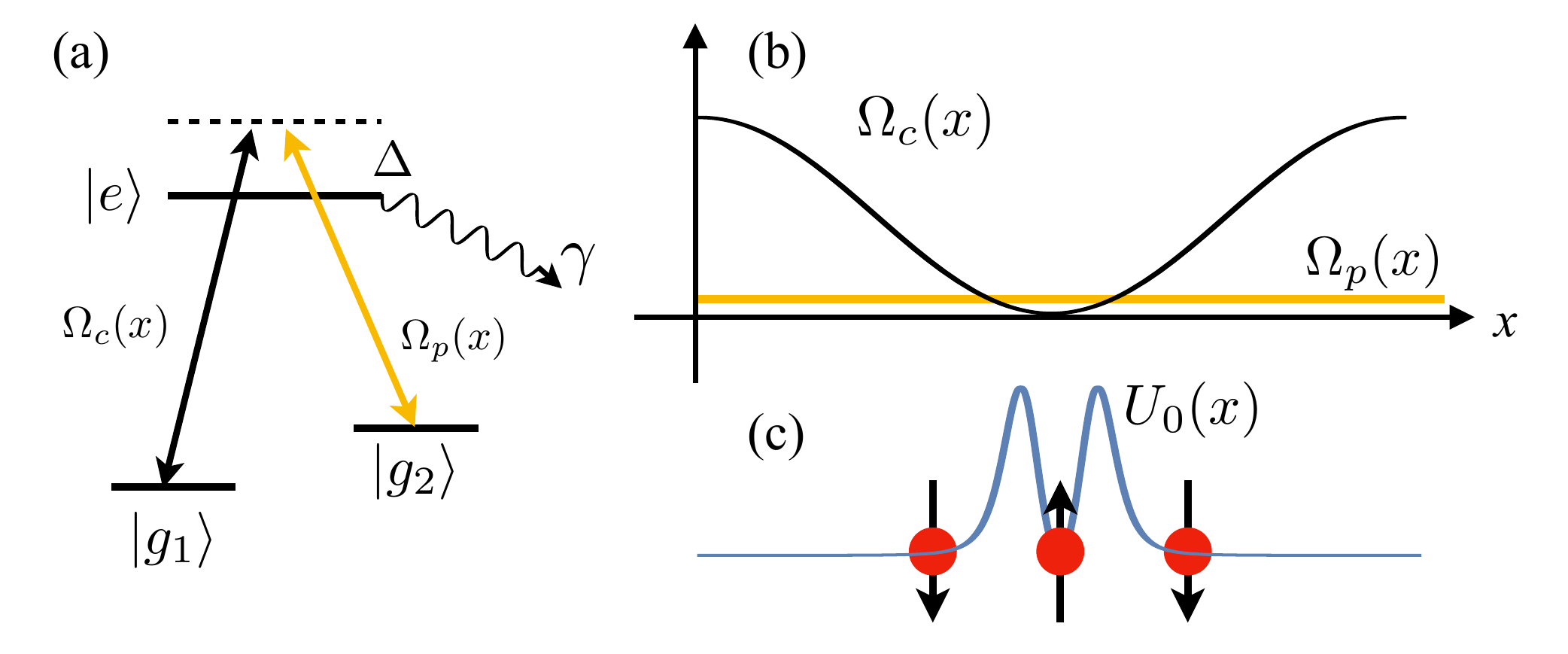}
\caption{Schematic of our scheme. (a) Atomic configuration and the coupling to two lasers. An example of (b) spatial dependences for $\Omega_c(x)$ and $\Omega_p(x)$ and (c) the corresponding non-adiabatic potential. The circular disks represent atoms and the arrows represent the population of the dark-state atoms.} 
\label{fig:scheme} 
\end{figure}

In this paper, we present a method to create multi-barrier non-adiabatic potentials with subwavelength spacing, realizing a new type of optical potential landscape without time-dependent modulations. Our idea is based on dark-state non-adiabatic potentials of three-level atoms in the $\Lambda$ configurations (Fig. \ref{fig:scheme} (a)) by controlling the spatial intensity variations on the driving lasers. We employ the Born-Oppenheimer (BO) approximation \cite{Dum:1996aa} to study the internal eigenstates of the atoms and non-adiabatic corrections from the atomic motions. We derive the general formula for the corrections of an arbitrary spatial intensity function. In particular, we show that double barriers and triple barriers with subwavelength spacings can be realized. 

Double-barrier potentials have been widely studied in solid-state systems, such as semiconductor heterostructures \cite{Goldman:1987aa,Petukhov:2002aa,Songmuang:2010aa}. They are important for understanding effects such as resonant tunneling \cite{Petukhov:2002aa} and quasi-bound states \cite{Nguyen:2009aa}. Our scheme offers a new platform to study double barriers in cold atoms in the subwavelength regime, where many-body atomic interactions can be strongly enhanced. As an example, we illustrate the formation of bound states of two and three atoms via magnetic dipolar interactions numerically. Our results show that the double-barrier potentials can support bound states of very long lifetime on the order of seconds.



\section{Dark-state trapping and non-adiabatic potentials}
We consider three-level atoms with a $\Lambda$ energy structure shown in Fig. \ref{fig:scheme} (a). The two ground states $\ket{g_1}$ and $\ket{g_2}$ are long-lived, which can be hyperfine-split metastable states or Zeeman sublevels \cite{Fleischhauer:2005aa}, and an excited state $\ket{e}$ has a spontaneous decay rate $\gamma$. The total Hamiltonian of the system includes the kinetic energy and the internal atomic interaction is given by \cite{lacki:2016aa}
$\mathcal{H}=\frac{p^2}{2m}+\mathcal{H}_{\text{in}}$, where
\begin{align}
\mathcal{H}_{\text{in}}=-\hbar\Delta\ket{e}\bra{e}+\left[\hbar\frac{\Omega_c(x)}{2}\ket{e}\bra{g_1}+\hbar\frac{\Omega_p(x)}{2}\ket{e}\bra{g_2}+\text{H.c.}\right].
\end{align}
Here $\Omega_c(x)$ ($\Omega_p(x)$) is the spatial-dependent Rabi frequency of the off-resonant coupling (probe) field for the transition between $\ket{e}$ and $\ket{g_1}$ ($\ket{g_2}$) with a detuning $\Delta$. These fields form a resonant Raman transition between the ground states \cite{SZ}. To study the non-adiabatic corrections, we will employ a similar procedure to that in Ref. \cite{Jendrzejewski:2016aa}. However, in contrast to the previous works \cite{lacki:2016aa,Jendrzejewski:2016aa, Wang:2018aa,subhankar2019floquet,lacki2019stroboscopic}, we consider both the coupling and the probe fields to be position dependent, which will lead to general expressions of the non-adiabatic corrections.

We are interested in the limit of slow atomic motions such that the energy of the external motion (the kinetic energy and the potential) is much smaller than the internal energy scale determined by $\mathcal{H}_{\text{in}}$. Within this limit, we can apply the BO approximation \cite{Dum:1996aa} to treat the internal Hamiltonian separately. The validity of this limit will be discussed later. We can diagonalize $\mathcal{H}_{\text{in}}$ and obtain the eigenstates to be
\begin{align}
&\ket{D(x)}=\frac{-\Omega_p(x)\ket{g_1}+\Omega_c(x)\ket{g_2}}{\Omega(x)},\nonumber\\
&\ket{B_{\pm}(x)}=\frac{\Omega_c(x)\ket{g_1}+\Omega_p(x)\ket{g_2}+\Omega_{\pm}(x)\ket{e}}{\sqrt{\Omega^2(x)+\Omega_{\pm}^2(x)}},
\end{align}
where the corresponding eigenenergies are $0$ and   $\hbar\Omega_{\pm}/2$, respectively. Here $\Omega(x)=\sqrt{\Omega^2_c(x)+\Omega^2_p(x)}$, $\Omega_{\pm}(x)=-\Delta\pm\sqrt{\Omega^2(x)+\Delta^2}$. Therefore, the atoms are trapped at the spatially varying dark state $\ket{D(x)}$ and are decoupled from the laser fields \cite{Agarwal:2006aa,Yavuz:2007aa,Gorshkov:2008aa, Kiffner:2008aa}. 

Since the momentum $p$ and the position $x$ do not commute, additional contributions arise when we diagonalize the kinetic energy using the position-dependent eigenbasis, which are termed as non-adiabatic corrections or geometric potentials \cite{lacki:2016aa, Jendrzejewski:2016aa}. To see this, we consider the unitary operator \cite{Jendrzejewski:2016aa}
$\mathcal{R}=\ket{D(x)}\bra{D_0}+\ket{B_{+}(x)}\bra{B_{+0}}+\ket{B_{-}(x)}\bra{B_{-0}},$
where $\ket{D_0}$ and $\ket{B_{\pm0}}$ are the eigenstates located at a fixed position $x_0$. By applying the transformation, we obtain
\begin{align}
\widetilde{\mathcal{H}}=\mathcal{R}^{\dagger}\mathcal{H}\mathcal{R}=\frac{\left(p-\mathcal{A}\right)^2}{2m}+\sum_{a=\pm}\hbar\frac{\Omega_{a}(x)}{2}\ket{B_{a0}}\bra{B_{a0}},
\end{align}
where the effective vector potential $\mathcal{A}\equiv\hbar\mathcal{R}^{\dagger}\partial_x\mathcal{R}=i\hbar\alpha^{\prime}(x)\sum_{a=\pm}N_{a}\left(\ket{B_{a0}}\bra{D_{0}}-\ket{D_{0}}\bra{B_{a0}}\right)+i\hbar\frac{\Omega^{\prime}(x)}{\Omega(x)} C\left(\ket{B_{-0}}\bra{B_{+0}}-\ket{B_{+0}}\bra{B_{-0}}\right)$
with $\alpha(x)=\arctan\left[\Omega_p(x)/\Omega_c(x)\right]$, $N_{\pm}=1/\sqrt{1+\Omega^2_{\pm}(x)/\Omega^2(x)}$ and $C=(\Delta\Omega/2)/(\Delta^2+\Omega^2)$. Therefore, we arrive at
\begin{align}
\widetilde{\mathcal{H}}&=\frac{p^2}{2m}+U_0(x)\ket{D_0}\bra{D_0}\nonumber\\
&+\sum_{a=\pm}\left(\hbar\frac{\Omega_{a}(x)}{2}+U_a(x)\right)\ket{B_{a0}}\bra{B_{a0}}+\mathcal{V},
\end{align}
where the non-adiabatic potential for the dark state is \cite{lacki:2016aa}
\begin{align}
U_0(x)=\frac{\hbar^2}{2m}\left[\alpha^{\prime}(x)\right]^2.
\label{eq:non-adia-potential}
\end{align}
It is the nonzero derivative of $\alpha(x)$ that gives rise to potential barriers (positive potential) for the dark state $\ket{D(x)}$. Thus the ratio of the two driving lasers, $f(x)\equiv\Omega_c(x)/\Omega_p(x)=\tan[\alpha(x)]$, plays an important role in engineering interesting spatial structures of subwavelength barriers. 
The non-adiabatic potentials for the bright states are 
$U_{a}=\hbar^2/(2m)\left[N_{a}^2\alpha^{\prime2}(x)+C^2\Omega^{\prime2}(x)/\Omega^2(x)\right]=N_{a}^2U_0(x)+4C^2U_1(x)\le U_0(x)+U_1(x)/4,$
where $U_1(x)=\hbar^2/(8m)\Omega^{\prime2}(x)/\Omega^2(x)$. 
The off-diagonal contribution that couples between the eigenstates is given by
\begin{align}
\label{eq:nac}
\mathcal{V}=-\frac{p\mathcal{A}}{2m}+U_{b}(x)\ket{B_{+0}}\bra{B_{-0}}+\sum_{a=\pm}U_{0a}(x)\ket{D_0}\bra{B_{a0}}+\text{h.c.},
\end{align}
where $U_b(x)=\hbar^2/(2m)N_{+}N_{-}\alpha^{\prime2}(x)$, and $U_{0a}=\pm \hbar^2/(2m)N_{\mp}C\alpha^{\prime}(x)\Omega^{\prime}(x)/\Omega(x)$. The coupling rates between the dark state and the bright states are $V_{\pm D}\equiv|\bra{B_{\pm0}}\mathcal{V}\ket{D_0}|/\hbar=|U_{0\pm}(x)|/\hbar$, which determine the loss rate of atoms out of the dark state. We show that $U_b(x)\le U_0(x)/2$ and $|U_{0\pm}(x)|\le \sqrt{U_0(x)U_1(x)}/2$. To ensure the validity of BO approximation, 
\begin{align}
U_{i}(x)/\hbar\ll | \Omega_{\pm} (x)| \quad (i=0, 1). 
\label{eq:inequality}
\end{align}
In particular, as shown numerically in Fig. \ref{fig:single} (b),  $U_{1}(x) \lesssim U_0(x)$ around the double barriers. For $|\Delta|\lesssim |\Omega(x)|$, we can reduce Eq.~\eqref{eq:inequality} to $U_{0}(x)\ll \hbar|\Omega(x)|$. Under this condition, the off-diagonal coupling $\mathcal{V}$ is far-detuned from the energy spacing between the eigenstates, therefore it can be treated as a perturbation. Then the excitation to the open channels can be estimated \cite{Jendrzejewski:2016aa, SZ} as $P_B\sim |\bra{B_{a0}}\mathcal{V}\ket{D_{0}}|^2/\Omega^2(x)= V^2_{\pm D}/\Omega^2(x)\ll1$. The effective loss rate of the dark state atoms is given by
$\gamma_d\sim\gamma V^2_{\pm D}/\Omega^2(x)$. A more rigorous result of the dark-state decay rate is calculated from the corrections to the dispersion of atoms in the Bloch bands \cite{lacki:2016aa}, which shows a similar relation to our perturbative result.






\section{Subwavelength optical potentials}
The non-adiabatic potential arises due to the rapidly spatial change of the internal dark state. Previous studies \cite{lacki:2016aa, Jendrzejewski:2016aa, Wang:2018aa, bienias2018coherent, subhankar2019floquet, lacki2019stroboscopic, Tsui} have focused on the situation when the ratio of the coupling laser to the driving laser to be an approximated linear function near certain values, i.e. $f(x)\approx kx/\epsilon$ for $|kx|\lesssim \epsilon$. The non-adiabatic potential for the dark state is given by \cite{Jendrzejewski:2016aa} $U_{0}(x)=\frac{\hbar^2k^2}{2m\epsilon^2}/[1+(kx/\epsilon)^2]^2,$
which shows a single potential barrier located at $x=0$ with $U_{0}(0)=\frac{\hbar^2k^2}{2m\epsilon^2}$ and a width $\Delta x\sim \epsilon/k$. In general,  $f(x)$ can be made periodic, e.g. $f(x)=\sin(kx)/\epsilon$ \cite{lacki:2016aa, Wang:2018aa}, so the separation between two barriers is $\lambda/2$.  Recent studies show that the separation can be reduced further via Floquet engineering \cite{subhankar2019floquet, lacki2019stroboscopic, Tsui}.

Here we consider a more general situation of laser intensities as
\begin{align}
f(x)=\frac{\Omega_c(x)}{\Omega_p(x)}=\frac{a+b\cos(kx)}{c+d\cos(kx+\phi)},
\end{align}
which can be formed by a combination of standing waves and propagating waves. Here $a,\ b,\ c,\ d,$ are the amplitude coefficients to be determined and $\phi$ is the phase control. 

\subsection{Double Barrier with subwavelength spacing}
\begin{figure}[t]
\leavevmode\includegraphics[width = 1. \columnwidth]{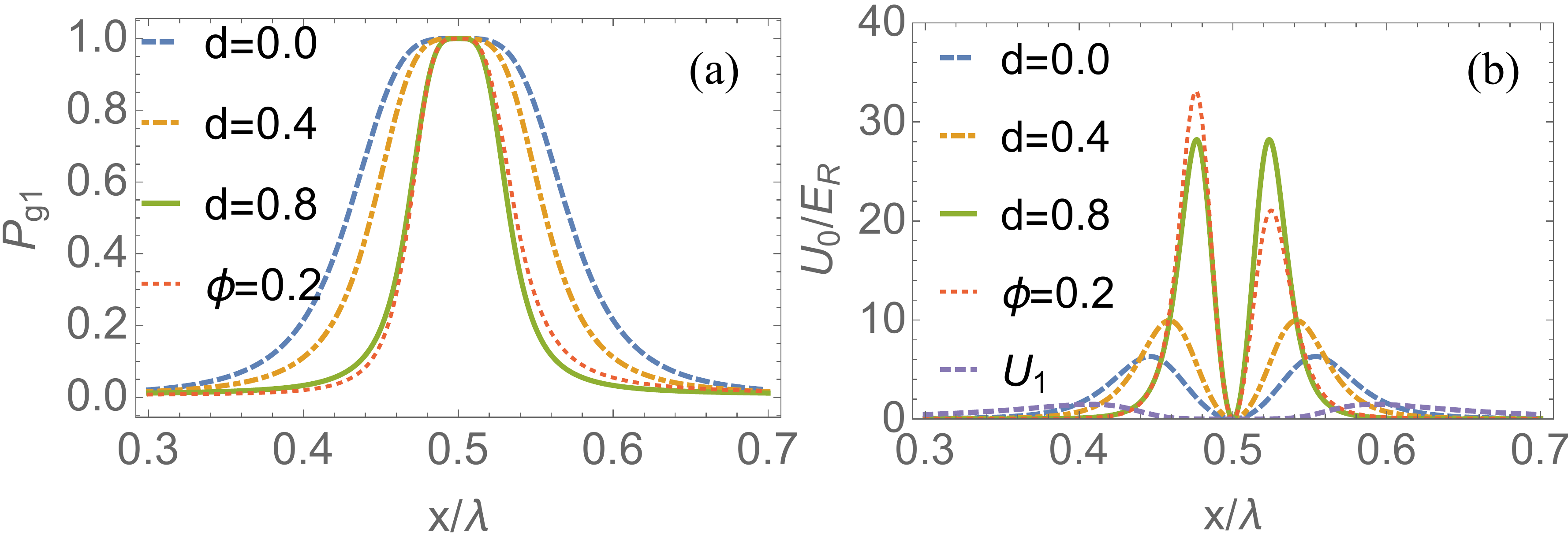}
\caption{(a) Dark-state population of $\ket{g_1}$ for double-barrier potentials as a function of $x/\lambda$ in one period, and (b) the corresponding non-adiabatic potentials of double barriers with subwavelength spacings for $d=0,~0.4~,0.8$ at $\phi=0$ and $d=0.8$ at $\phi=0.2$ with $\epsilon=1/10$. A component of the non-adiabatic potential for the bright states ($U_1$, the dashed curve) is plotted in (b) for $d=0$ and $\phi=0$.} 
\label{fig:single} 
\end{figure} 

We study spatial engineering of the Rabi frequencies to realize non-adiabatic multiple barriers with subwavelength spacing. The basic idea is to design a spatial function $f(x)$ such that its derivative is zero at $x=x_{\min}$, but it quickly reaches its maximum in a subwavelength region away from $x_{\min}$.

We consider the situation for generating double barriers using
\begin{align}
f(x)=\frac{1}{\epsilon}\frac{1+\cos(kx)}{1+d\cos(kx+\phi)},
\label{eq:f-single}
\end{align}
where $d<1$. Here we take $\Omega_c(x)=\Omega_0\left[1+\cos(kx)\right]$ and $\Omega_p(x)=\Omega_0\epsilon\left[1+d\cos(kx)\right]$. 

At $\phi=0$, we find that the potential features are prominent at  $kx\sim \pi$ and we arrive at
\begin{align}
\alpha^{\prime}(x)\approx\frac{k}{\epsilon(1-d)}\frac{k\delta x}{1+\left[\frac{k^2\delta x^2}{2\epsilon(1-d)}\right]^2},\label{eq:a-single}
\end{align}
where $\delta x = x -\pi/k$. The spatial function of the non-adiabatic potential $U_0(x)$ can be obtained from Eq. \eqref{eq:non-adia-potential} by substituting the above relation of $\alpha^{\prime}(x)$, from which we obtain three important features:\\ 
\noindent (1) $U_0(x_{\min})=0$ with $x_{\min}=\pi/k$; \\
\noindent (2) two barrier peaks located around $x_{\max}=\pi/k\pm (4/3)^{1/4}\frac{\sqrt{\epsilon(1-d)}}{k}$ with $U_0(x_{\max})=\frac{\hbar^2k^2}{2m}\frac{\sqrt{27}}{8\epsilon(1-d)}$; \\
\noindent (3) we get a non-adiabatic potential well with a subwavelength width at half maximum to be $\Delta x\approx 0.2\sqrt{\epsilon(1-d)}\lambda$.

Furthermore, we study spatial variation on the population of one of the ground states, e.g.  $P_{g_1}(x)=\frac{1}{1+f^2(x)}$ . The non-adiabatic potential $U_0(x)\propto (\partial P_{g_1}(x)/\partial x)^2$, meaning that it is due to the rapid change of the population as a function of position such that the atoms can not adjust its internal state adiabatically. In particular, $\partial P_{g_1}(x)/\partial x\propto f^{\prime}(x)=0$ at $x=k/\pi$, which corresponds to the dip in the non-adiabatic potential. 

We plot the numerical results of the ground state population $P_{g_1}$ and the double-barrier potential $U_0(x)$ for $d=0,~0.4,~0.8$ at $\phi=0$ and $d=0.8$ at $\phi=0.2$ in Fig. \ref{fig:single} (a) and (b), respectively, using the full expression derived from Eq. \eqref{eq:f-single}. The aforementioned features agree well with the numerical results. The example of $\phi=0.2$ shows the ability to tune the heights of the potential barriers from symmetric into asymmetric.

In addition, we note that for any function  $f(x)\approx (x-x_{\min})^n/\epsilon$ for $n >1$ (intergers), the corresponding geometric potential can be a double barrier  centered at $x=x_{\min}$ since $f^{\prime}(x_{\min})=0$. Here, we have provided a simple realization of non-adiabatic potential double barriers, making them drastically different from those in the case of $f(x)\approx x/\epsilon$ \cite{lacki:2016aa, Jendrzejewski:2016aa, Wang:2018aa, bienias2018coherent}. Our method does not require lattice modulation \cite{ subhankar2019floquet, lacki2019stroboscopic, Tsui} or atomic levels with multi-$\Lambda$ configurations \cite{Kiffner:2008aa, lacki:2016aa}.




\subsection{Multiple Barrier with subwavelength spacing}
\begin{figure}[t]
\leavevmode\includegraphics[width =  \columnwidth]{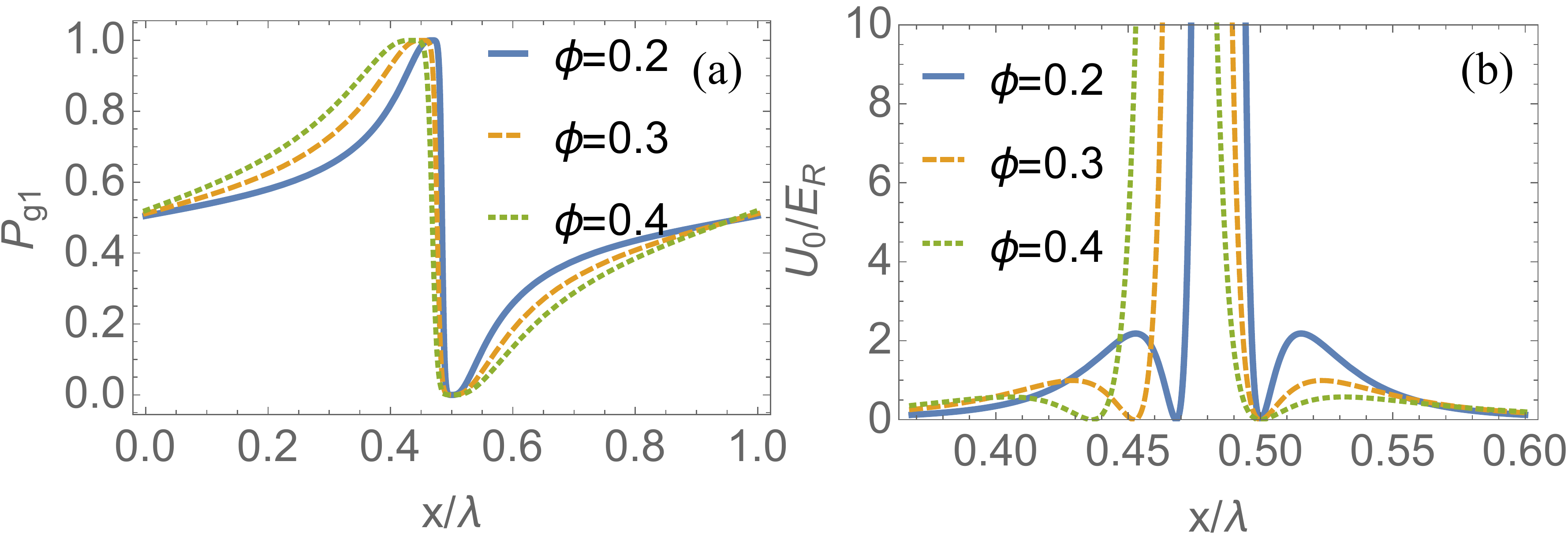}
\caption{(a) Dark-state population of $\ket{g_1}$ for creating triple-barrier potentials as a function of $x/\lambda$, and (b) the corresponding non-adiabatic potentials of triple barriers with subwavelength spacings for different values of $\phi$. The inset magnifies the dashed-box area to show the triple-barrier feature.} 
\label{fig:double} 
\end{figure} 

As another example, we would like to show a three-peak non-adiabatic potential.  To see this, we consider the spatial function 
$f(x)=\frac{1+\cos(kx)}{1+\cos(kx+\phi)},$
where $\Omega_c(x)=\Omega_0\left[1+\cos(kx)\right]$ and $\Omega_p(x)=\Omega_0\left[1+\cos(kx+\phi)\right]$. Here the only control parameter is the phase $\phi$. We derive
\begin{align}
\alpha^{\prime}(x)=k\frac{\sin\left( kx+\phi\right)-\sin kx +\sin \phi}{(1+\cos kx)^2+(1+\cos \left( kx+\phi\right))^2}.
\end{align}
By examining the properties of $\alpha^{\prime}(x)$, we find the spatial features of $U_0(x)$ as follows:\\
(1) A central peak at $x_{\text{c}}=\frac{\pi}{k}-\frac{\phi}{2k}$ with $U_0(x_{\text{c}})=\frac{8\hbar^2k^2}{m\phi^2}$;\\
(2) Two dips at $x_{\min}=x_{\text{c}}\pm\frac{\phi}{2k}$ with $U_0(x_{\min})=0$ ; \\
(3) Two lower peaks at maximum at $x_{\max}=x_{\text{c}}\pm\frac{\phi}{k}$ with $U_0(x_{\max})=\frac{9\hbar^2k^2}{200m\phi^2}$.

We plot both the ground-state population $P_{g_1}$ and the geometric potential $U_0(x)$ as a function of $x$ for different values of $\phi$ in Fig. \ref{fig:double}. Two flat regions in the population correspond to the dips in $U_0(x)$ and the large slope corresponds to the central peak in $U_0(x)$.


We note that the double-well potential looks similar to the plot in the right-upper panel in Fig. 3 (b) in Ref. \cite{Wang:2018aa}, but they are of different natures. Here the potential is a characteristic of the spatially varying Rabi frequencies and the state is complete dark, while the one in Ref.  \cite{Wang:2018aa} is due to the off-resonant Raman lasers so that the resulting potential is not complete dark.


The non-adiabatic multiple barriers are important in two ways. First, they provide a new subwavelength potential landscape. Quasi-bound states and resonant tunnelings of multi-barrier potentials can be studied using Wentzel-Kramers-Brillouin (WKB) approximation \cite{Modinos:1969aa, Dutt:2010aa}. Second, the subwavelength features of the multiple barriers may permit the study of enhanced dipole-dipole interactions between atoms. 


\section{Application: bound states \label{sec:mbp}}

\begin{figure}[t]
\leavevmode\includegraphics[width = .75 \columnwidth]{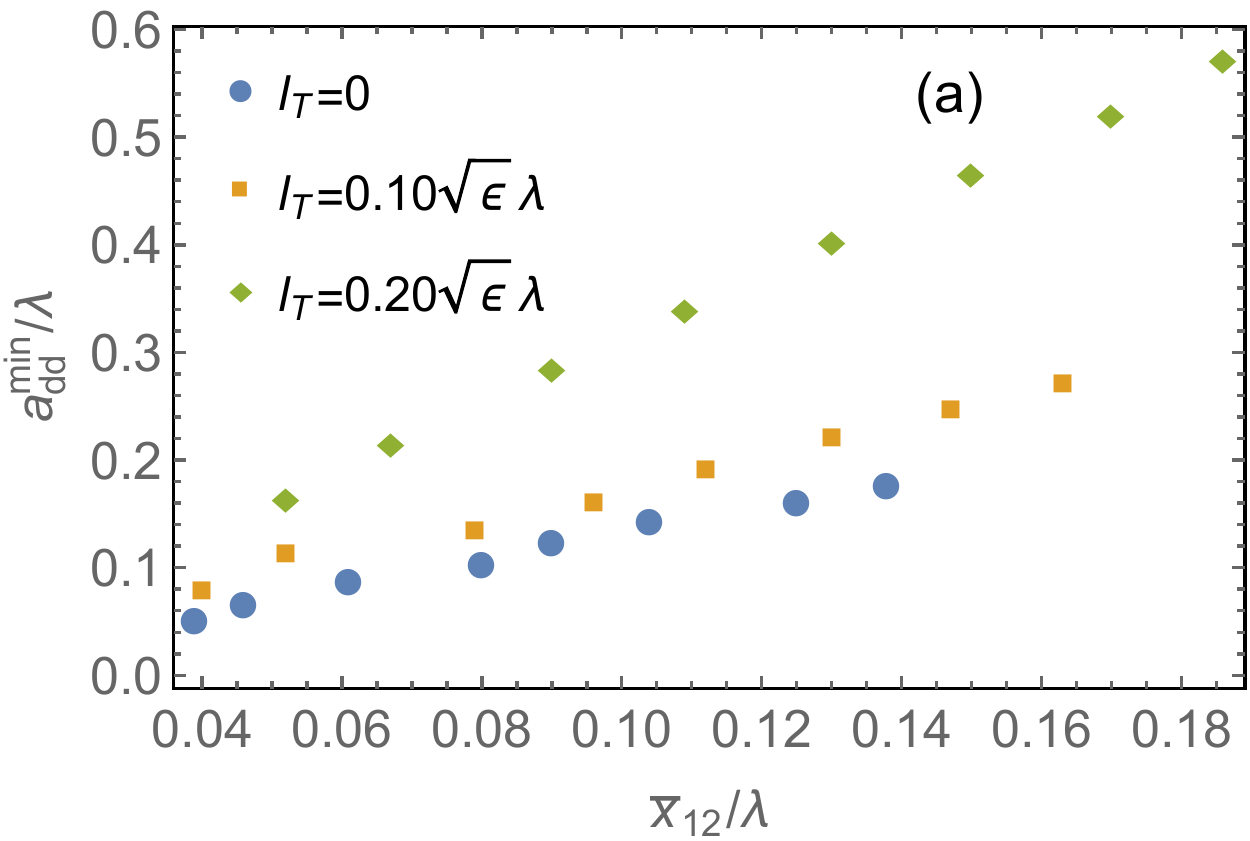}
\leavevmode\includegraphics[width =  \columnwidth]{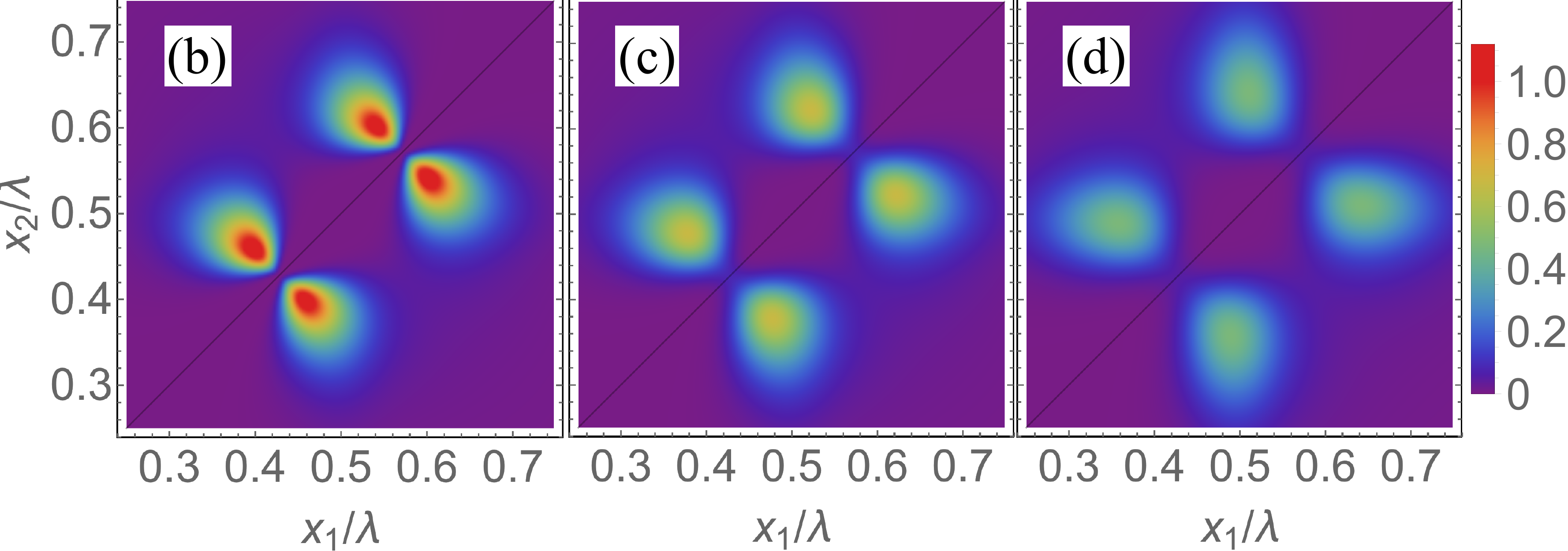}
\caption{(a) The minimum values of the dipolar length $a^{\min}_{dd}$ versus the average atomic distance $\bar{x}_{12}$ for different strengths of transverse confinement $l_T$. For each value of confinement, $\bar{x}_{12}$ is evaluated numerically from the corresponding  wave function $\psi(x_1,x_2)$ solved using $a^{\min}_{dd}$ with $\epsilon=1/120,~1/70,~1/40,~1/28,~1/20,~1/15,~1/12,~1/10$ at $d=0$. Density plots of $|\psi(x_1,x_2)|^2$ for (b) $l_T=0$, (c) $l_T=0.10  \sqrt{\epsilon}\lambda$, (d) $l_T=0.20  \sqrt{\epsilon}\lambda$ using $a^{\min}_{dd}$ with $\epsilon=1/10$.}
\label{fig:add} 
\end{figure} 

We now discuss a study of many-body physics permitted by the non-adiabatic potential barriers. The spatial-varying dark state may allow the atoms to have spatially dependent magnetic moments \cite{Bloch:2008aa} in the subwavelength regime, leading to strong magnetic dipole-dipole interactions. For concreteness, we assume the magnetic moments are aligned perpendicular to $x-$axis that the atoms are located at. (The situation on transverse position distribution will be discussed later.) The strength of the magnetic moment of an atom is taken to be $\mu(x)=\mu_m\left[2P_{g_1}(x)-1\right]$, which depends on the state (spin) of the atom, i.e., $P_{g_1}(x)$. For the situation of two atoms, the total Hamiltonian is given by \cite{lacki:2016aa}
\begin{align}
\label{eq:MB}
\mathcal{H}_{MB}=\frac{p_1^2}{2m}+\frac{p_2^2}{2m} + U_0(x_1)+U_0(x_2)+\frac{\mu_0\mu(x_1)\mu(x_2)}{4\pi|x_1-x_2|^3},
\end{align}
where $\mu_0$ is the vacuum permeability, and $x_i$ and $p_i$ are the position and the momentum of the $i$th atom.


In the case of the double barriers,  the atoms repel each other through the magnetic dipolar interaction if they both stay in the well.
Thus a bound state can not be formed. Instead, a bound state can be formed if only one atom sits inside the well. The condition for the bound state is $E_{\min}+\frac{\mu_0\mu(x_1)\mu(x_2)}{4\pi|x_1-x_2|^3}<0$, where $E_{\min}$ is the minimum energy of the atom inside the potential well and the energy of the other atom outside the well is neglected. By considering the size of the potential well and using the uncertainty principle, we estimate $E_{\min}\approx \hbar^2k^2/\left[2m\epsilon(1-d)\right]$. Then we estimate the minimum magnetic moment in order to form a bound state, and we find that the required dipolar length $a_{dd}\equiv\mu_0\mu^2m/\left(12\pi\hbar^2\right)$ \cite{Lahaye:2009aa} is on the order of the typical subwavelength spacing, i.e., $a^{\min}_{dd}\sim 0.2\sqrt{\epsilon(1-d)}\lambda\sim\Delta x$. 

A more rigorous determination of $a^{\min}_{dd}$ can be obtained by numerically solving the Sch\"{o}rdinger equation \eqref{eq:MB} to have at least one negative eigenvalue. In practice, the dipole-dipole interaction would be modified due to the transverse distribution, and so does $a^{\min}_{dd}$. The dipolar potential in 3D is given by $V_{\text{3D}}=\mu_0\mu(x_1)\mu(x_2)(r^2-3z_{12})/(4\pi r^5)$ \cite{Buchler07}. Here $r$ is the atomic distance in 3D, and $z_{12}$ is the atomic distance along the $z-$axis, where the dipoles are assumed to be oriented. We assume that each atom is strongly confined in the transverse direction, where its distribution probability for the $j$th atom is given by $P(y_j,z_j)=\exp[-(y_j^2+z_j^2)/l_T^2]/(4\pi l_T^2)$ with $l_T$ the strength of transverse confinement. We first find the effective dipole-dipole interaction by integrating the transverse variables in $V_{\text{3D}}$ using $P(y_j,z_j)$, and then solve Eq. \eqref{eq:MB} using the effective interaction. 

We study the dependence of $a^{\min}_{dd}$ shown in Fig. \ref{fig:add} (a). In the ideal case when $l_T=0$, we search $a^{\min}_{dd}$ for different values of $\epsilon$ (circular points). Using the critical values of $a_{dd}$, we calculate the average atomic distance along the $x-$axis $\bar{x}_{12}$ from the wave function $\psi(x_1,x_2)$. We find that $a^{\min}_{dd}$ increases linearly with $\bar{x}_{12}$ as we change $\epsilon$, i.e., $a^{\min}_{dd}\approx 1.2 \bar{x}_{12}\approx 0.55 \sqrt{\epsilon}\lambda$, which is close to the value obtained in the analytical method when $d=0$. For $l_T= 0.10 \sqrt{\epsilon}\lambda\ll \bar{x}_{12}$, we find both $\bar{x}_{12}$ and $a_{dd}^{\min}$ increase slightly where $a^{\min}_{dd}\approx 1.5 \bar{x}_{12}$ (square points). For $l_T= 0.20  \sqrt{\epsilon}\lambda\approx \bar{x}_{12}/3$, $a^{\min}_{dd}$ is increased to about $3.0 \bar{x}_{12}$ (diamond points). 

We plot the bound-state probability distributions $|\psi(x_1,x_2)|^2$ using $a^{\min}_{dd}$ at $\epsilon=1/10$ for $l_T=0,\ 0.10 \sqrt{\epsilon}\lambda, \ 0.20 \sqrt{\epsilon}\lambda$ in Fig. \ref{fig:add} (b)-(d). We observe that there is a high probability to find one atom located inside the potential well while the other outside, so the magnetic moments tend to be close to its maximum magnitude $\mu_m$. There is almost zero probability to find atoms sitting close to the domain walls ($x_j\approx 4.25$ and $5.75$) where $\mu(x)$ changes signs \cite{lacki:2016aa}, so the contact interaction due to $s$-wave scattering \cite{Lahaye:2009aa} can be neglected. 

\begin{figure}[t]
\leavevmode\includegraphics[width = .85 \columnwidth]{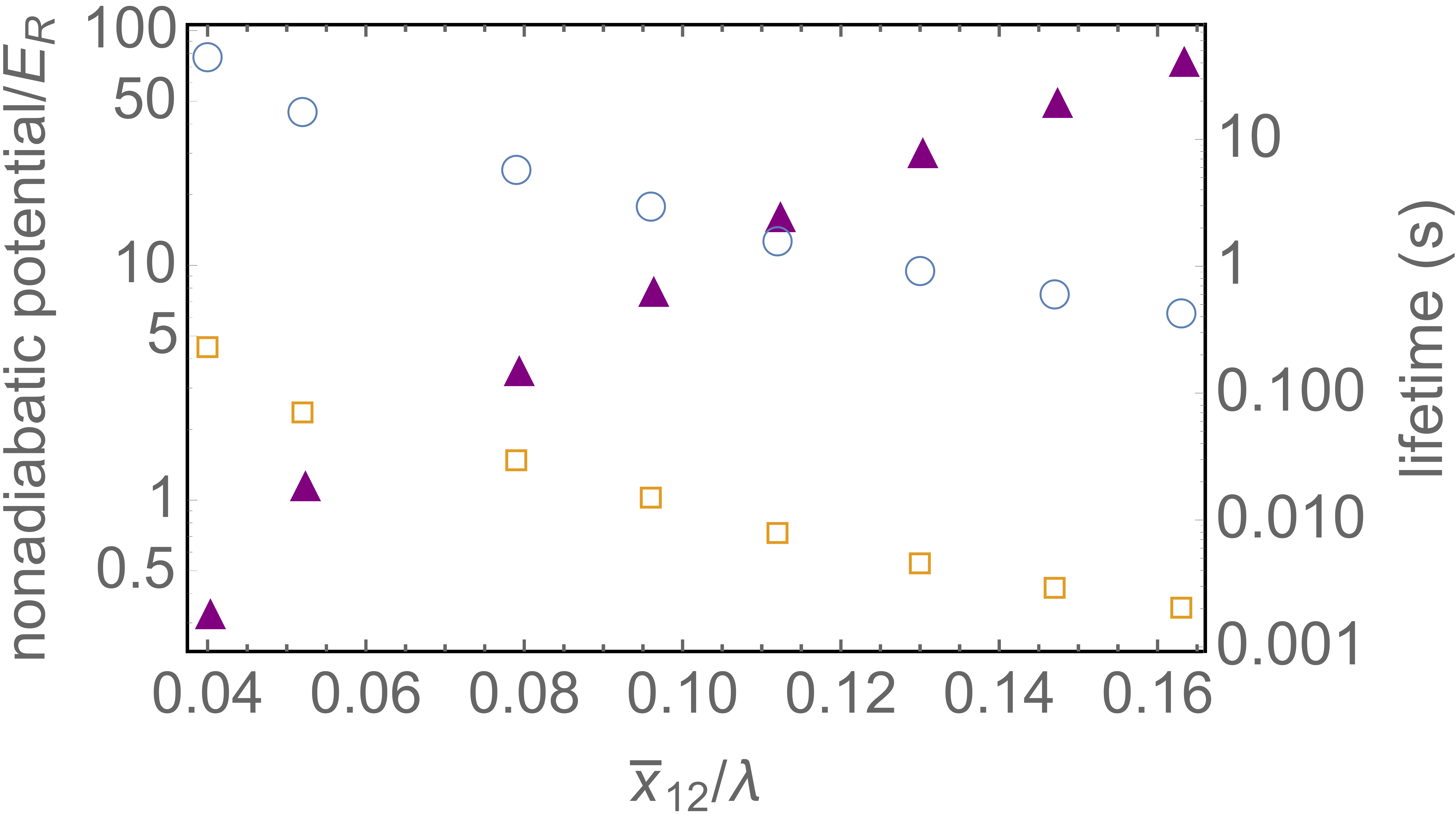}
\caption{The average non-adiabatic potential of a bound-state atom (open squares), the maximum potential of the double barriers (circles), and the lifetime of the bound states (triangles), versus the mean atomic distance with the same set of $\epsilon$ in Fig. \ref{fig:add} (a) for $l_T=0.1\sqrt{\epsilon}\lambda$. (see text for details.) }
\label{fig:pa} 
\end{figure}

We also investigate the nonadiabatic potential experienced by the bound-state atoms. Due to the double-barrier potential, the average off-diagonal coupling rate can be very small, which is given by $\int V_{\pm D}(x_1)|\psi(x_1,x_2)|^2dx_1dx_2\le \int \sqrt{U_0(x_1)U_1(x_2)}/(2\hbar)\psi(x_1,x_2)|^2dx_1dx_2\equiv \bar{U}_{\text{off}}/\hbar$. In Fig. \ref{fig:pa} (empty squares), we plot this upper-bound of the average off-diagonal coupling potential $\bar{U}_{\text{off}}$ versus the mean atomic distance as $\epsilon$ varies. The results show that $\bar{U}_{\text{off}}\sim E_R$, which is two orders of magnitude smaller than the peak nonadiabatic potential $U_0(x_{\max})$ in Fig. \ref{fig:pa} (circles). The small off-diagonal coupling potential can support very long lifetime of the bound states. We estimate the lifetime of the bound states as $\tau=1/\bar{\gamma}_d$, where $\bar{\gamma}_d=\gamma\int U_0(x_1)U_1(x_2)/(2\hbar\Omega(x_1))^2|\psi(x_1,x_2)|^2dx_1dx_2$. Using the parameters in Sec. \ref{V}, we show that lifetime on the order of seconds can be realized at tens of nanometers atomic distance.

Moreover, the double-barrier potential can lead to bound states of three atoms known as trimers.  Our trimer has one atom inside the well and two outside on each side of the barrier. Similarly, bound states can be formed in the case of the triple potential barriers. The atoms attract each other if they are separated by the center barrier such that the dipoles are opposite and the magnitude is about $\mu_m$ (Fig. \ref{fig:double}). With periodic potentials, our multiple barriers can be interesting for the study of many-body physics in band structures, which is out of the scope of this work. 



\section{Experimental Implementations\label{V}}
The non-adiabatic potential multiple barriers require the $\Lambda$ atomic configuration and spatial control on the Rabi frequencies. The $\Lambda$ atomic configurations have been frequently used in electromagnetic-induced transparency \cite{Fleischhauer:2005aa} and coherent population transfer \cite{Bergmann:1998aa}. The first experiment on non-adiabatic potentials has been performed using an ultracold $^{171}$Yb gas \cite{Wang:2018aa} with hyperfine-split ground states.

The spatial dependence on the Rabi frequencies $\Omega_c(x)$ and $\Omega_p(x)$ is of the form $a+b\cos(kx)$, which can be realized by superposing a standing-wave and a propagating wave derived from the same laser. Thus, the spatial function $f(x)$ is insensitive to the laser intensity fluctuation. As the perfect spatial function may be challenge in experiment, a double barrier can be more easily demonstrated with an approximated function $f(x)=(x-x_{\min})^2/\epsilon$ by shaping the coupling and the probe lasers.

We can determine the minimum barrier spacing from $U_{0}(x)\ll \hbar|\Omega(x)| $. The minimum value of the Rabi frequency is $\Omega(x)=\sqrt{\Omega_c^2(x)+\Omega_p^2(x)}\approx \Omega_0\epsilon (1-d)$ for the double-barrier potentials, and $\Omega(x)\approx \Omega_0\left[1-\cos^2(\phi)\right]\approx \Omega_0\phi^2/2$ for the triple-barrier potentials. Considering $\Omega_0=2\pi\times 100$ MHz, $\lambda=532$ nm, and $U_{0}(x)\le \hbar|\Omega(x)|/5$ for $^{171}$Yb atoms, we obtain the minimum barrier spacing to be $13$ nm and $24$ nm for the double-barrier potentials and the triple-barrier potentials, respectively. The corresponding minimum energy inside the double barriers is $E_{\min}/\hbar\approx 2\pi\times 280$ kHz. We estimate the maximum excitation probability is $P_B\approx 4\%$. Taking $\gamma=2\pi\times182$ kHz \cite{Wang:2018aa}, we find the scattering rate on the dark states through the open channel is about $2\pi\times7$ kHz. In the above analysis, we assumed $|\Delta|\lesssim|\Omega(x)|$, however, for larger single-photon detunings, the laser intensities need to be stronger in order to get the same subwavelength feature without breaking the BO approximations.

\section{Conclusion}
In this work, we presented a method of creating non-adiabatic potentials of multiple barriers separated at tens of nanometers via spatial engineering of the laser intensities. The ability of spatial control could potentially open a new direction to engineer interesting subwavelength potential landscapes.  We studied several concrete examples on realizing subwavelength multiple barriers and their application for bound states. Interestingly, we show that double-barrier potentials can support bound states of very long lifetimes.

The multiple barriers can be a new platform for the study of many-body interactions in cold atoms enhanced by the subwavelength features. This scheme may also allow further studies on trapping atoms without the conventional optical dipole potentials and super-resolution quantum microscopy \cite{McDonald:2019aa, Subhankar:2019aa}.
\begin{acknowledgements}
This research is supported by a grant from King Abdulaziz City for Science and Technology (KACST)
\end{acknowledgements}




\bibliography{subwavelength}

\end{document}